\documentclass[twocolumn,aps,amsmath,amssymb,floatfix,superscriptaddress,prb,longbibliography]{revtex4-2}

\usepackage{graphicx}
\usepackage{dcolumn}
\usepackage{bm}
\usepackage{xspace}
\usepackage[colorlinks]{hyperref}
\usepackage{color}
\usepackage[flushleft]{threeparttable}
\usepackage{siunitx}
\def\MMO{Mn$_2$Mo$_3$O$_8$\xspace}
\def\new{\color{black}}

\begin{document}
	\title{Magnetic Interactions in the Polar Ferrimagnet with a Bipartite Structure}
	\author{Junbo Liao}
	\author{Zhentao~Huang}
    \author{Bo~Zhang}
	\author{Yanyan Shangguan}
	\author{Shufan~Cheng}
	\author{Hao~Xu}
	\author{Zihang~Song}
	\author{Shuai~Dong}
	\affiliation{National Laboratory of Solid State Microstructures and Department of Physics, Nanjing University, Nanjing 210093, China}
	\author{Devashibhai Adrojia}
	\affiliation{ISIS Facility, Rutherford Appleton Laboratory, Chilton, Didcot, Oxon OX11 0QX, United Kingdom}
	\affiliation{Highly Correlated Matter Research Group, Physics Department, University of Johannesburg, P.O. Box 524, Auckland Park 2006, South Africa}
	\author{Song~Bao}
	\email{songbao@nju.edu.cn}
	\author{Jinsheng~Wen}
	\email{jwen@nju.edu.cn}
	\affiliation{National Laboratory of Solid State Microstructures and Department of Physics, Nanjing University, Nanjing 210093, China}
	\affiliation{Collaborative Innovation Center of Advanced Microstructures and Jiangsu Physical Science Research Center, Nanjing University, Nanjing 210093, China}

\begin{abstract}
The polar magnets A$_2$Mo$_3$O$_8$ (A=Fe, Mn, Co, and Ni) feature a bipartite structure, where the magnetic A$^{2+}$ ions occupy two different sites with octahedral and tetrahedral oxygen coordinations. This bipartite structure provides a platform for the emergence of nontrivial magnetoelectric (ME) effects and intriguing excitation behaviors, and thus creates significant research interest. In this study, we conduct inelastic neutron scattering measurements on single crystals of Mn$_2$Mo$_3$O$_8$, an L-type ferrimagnet in the A$_2$Mo$_3$O$_8$ family, to investigate its spin dynamics. The obtained magnetic excitation spectra reveal two distinct magnon dispersions corresponding to the octahedral and tetrahedral spins in \MMO. These magnon bands can be well described by a spin Hamiltonian including Heisenberg and single-ion anisotropy terms. Employing our effective spin model, we successfully reproduce the unusual temperature dependence of the L-type ferrimagnetic susceptibility through self-consistent mean-field theory. This research reveals the significance of the bipartite structure in determining the excitation properties of the polar magnets $\rm{A_{2}Mo_{3}O_{8}}$ and provides valuable insights into the spin dynamics of L-type ferrimagnets.
\end{abstract}
\maketitle

\section{Introduction}
As a static property, structure profoundly influences dynamic aspects like transport and excitations through interactions in materials, and thus exploring exotic phenomena arising from novel structures has become a main stream in condensed matter physics~\cite{PhysRevLett.123.040601, PhysRevLett.129.207201, yan2022triangular, wang2023quantum,ghimire2020topology, RevModPhys.83.407, RevModPhys.88.041002}. From the perspective of spin dynamics, there are at least three classes of intriguing physics driven by structure: \romannumeral1. Topological magnon excitations protected by specific spin structural   symmetries~\cite{PhysRevX.8.041028,bao2018discovery,zhu2021topological,annurev:/content/journals/10.1146/annurev-conmatphys-031620-104715}; \romannumeral2. fractional excitations in a quantum spin liquids with geometrically frustrated structures~\cite{RevModPhys.89.025003,Savary_2017,npjqm4_12}; \romannumeral3. unique interactions resulting from structure, such as the Dzyaloshinskii-Moriya (DM) interaction~\cite{PhysRev.120.91, PhysRevB.99.214420, PhysRevLett.122.257202}, which induces novel magnetic states~\cite{PhysRevB.99.214420, RevModPhys.95.025001}.

\begin{figure}[tb]
	\centerline{\includegraphics[width=8.cm]{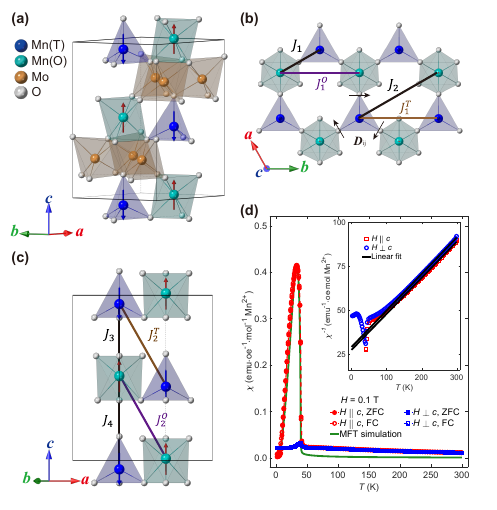}}
	\caption{(a) Schematic structures of the primitive cell of \MMO. Arrows indicate the magnetic moments on Mn$^{2+}$ ions. Mn(T) and Mn(O) denote the Mn ions at the tetrahedral and octahedral sites, respectively. (b) and (c) The structure of the Mn-O polyhedrons in the (0,~0,~1) and (0,~1,~0) planes, respectively. Solid lines illustrate the magnetic exchange interaction paths. Arrows in (b) denote the direction of DM vectors $\bm D_{ij}$ (d) Temperature dependence of the magnetic susceptibility $\chi$ with an external field of 0.1~T applied parallel and perpendicular to the $c$-axis. Note that the data show no difference between the zero-field-cooling and field-cooling conditions. The solid line is the calculated net spin as shown in Fig.~\ref{fig4} using the mean-field theory. The inset shows the temperature dependence of the inverse magnetic susceptibility $\chi^{-1}$. Solid lines in the inset denote the Curie-Weiss fits of $\chi^{-1}$.
		\label{fig1}}
\end{figure}

The polar magnets A$_{2}$Mo$_{3}$O$_{8}$ (A= Fe, Mn, Co, and Ni), well-known for their multiferroic properties~\cite{PhysRevX.5.031034,Wang2015,PhysRevLett.131.136701,PhysRevB.100.134112,PhysRevB.95.045142,PhysRevB.103.014112}, possess a bipartite polar structure. The crystal structure of these compounds belongs to a polar space group $P6_{3}mc$ (No.186)~\cite{bertrand1975etude,STROBEL1983329,STROBEL1982242,MCALISTER1983340}, which breaks space-inversion symmetry and allows for crystallographic polarity along the $c$ axis. It consists of alternating $\rm Mo$-$\rm O$ layers and $\rm A$-$\rm O$ layers stacking along the $c$ direction, as illustrated in Fig.~\ref{fig1}(a). Within the $\rm A$-$\rm O$ layer, $\rm A^{2+}$ ions adopt two distinct oxygen coordination environments with $\rm AO_{6}$ octahedra and $\rm AO_{4}$ tetrahedra. These $\rm AO$ polyhedra collaboratively form a bipartite honeycomb structure in the $a$-$b$ plane, as depicted in Fig.~\ref{fig1}(b). Below the magnetic ordering temperature, simultaneous breaking of time-reversal
and space-inversion symmetry occurs, which results in nontrivial magnetoelectric effects observed in the A$_{2}$Mo$_{3}$O$_{8}$ family~\cite{PhysRevX.5.031034,Wang2015,PhysRevLett.131.136701,PhysRevB.100.134112,PhysRevB.95.045142,PhysRevB.103.014112}. In addition, the different magnetic A$^{2+}$ ions in this bipartite polar structure, with varying electron filling of the 3$d$ shell, spin-orbital coupling, and single-ion anisotropy, create an intricate interplay that gives rise to abundant physical phenomena and excitation spectra in the A$_{2}$Mo$_{3}$O$_{8}$ family~\cite{ideue2017giant,bao2023direct,PhysRevB.95.020405,wu2023fluctuation, reschke2022confirming,PhysRevB.108.024431, gao2023diffusive,PhysRevB.102.094307,PhysRevB.102.174407,PhysRevB.102.115139,PhysRevLett.119.077206}.

For instance, in $\rm Fe_{2}Mo_{3}O_{8}$, a thermal transport study revealed a giant thermal Hall effect, which was considered as evidence of unconventional coupling between magnetism and phonons~\cite{ideue2017giant}. This finding was confirmed by a recent inelastic neutron scattering (INS) study~\cite{bao2023direct}, where the magnon-phonon hybrid excitations, referred to as magnon polarons, were directly observed. Additionally, it was also suggested that the DM interaction, driven by the absence of inversion center in the bipartite structure~\cite{PhysRev.120.91}, plays a significant role in the presence of the magnon-phonon coupling and the consequent emergence of magnon polarons~\cite{bao2023direct}. Magneto-Raman spectroscopy studies have shown the existence of fluctuation-enhanced phonon magnetic moments~\cite{wu2023fluctuation}, highlighting the role of many-body correlations and fluctuations in this compound. Regarding $\rm Co_{2}Mo_{3}O_{8}$, a THz spectroscopy research has confirmed the trilinear form of the optical magnetoelectric effect~\cite{reschke2022confirming}, implying the presence of electromagnon excitations. Meanwhile, the spin-orbital coupling of the $d^{7}$ Co$^{2+}$ ions in octahedral crystal-field enviroments gives rise to a $J_{\rm eff}=1/2$ effective-spin doublet~\cite{PhysRevB.97.014407,PhysRevB.97.014408}. Therefore, substituting the tetrahedral Co$^{2+}$ ions with nonmagnetic Zn$^{2+}$ ions results in a $J_{\rm eff}=1/2$ triangular-lattice antiferromagnet CoZnMo$_3$O$_8$~\cite{PhysRevB.108.024431}, providing a possible way to realize quantum-spin-liquid state. In the case of $\rm Ni_{2}Mo_{3}O_{8}$, crystal electric field splitting of $\rm Ni^{2+}$ ions was expected to lead to nonmagnetic singlet ground states~\cite{gao2023diffusive}. However, the interplay between exchange interactions and different crystal electric field gaps in octahedral $\rm Ni^{2+}$ and tetrahedral $\rm Ni^{2+}$ induces nontrivial antiferromagnetism~\cite{PhysRevMaterials.3.014410,PhysRevB.103.014112} and spin exciton excitations~\cite{gao2023diffusive}. 

Despite significant variations in the excitation behaviors with different magnetic ions in the A$_{2}$Mo$_{3}$O$_{8}$ family, it is evident that the bipartite structure common in this family consistently plays a fundamental role in determining these exotic properties. \MMO is another isostructural compound within this polar magnetic family. It exhibits an unusual L-type ferrimagnetism and complex magnetic interactions in the bipartite lattice structure~\cite{MCALISTER1983340,PhysRevB.95.045142,10.1063/1.333657}. In this Letter, by performing INS measurements on single crystals of \MMO, we map out the entire spin excitation spectra. Utilizing linear spin-wave theory, we determine the effective spin model. The extracted exchange parameters reveal distinct dynamic properties of spins at octahedral and tetrahedral sites. With the obtained parameters, we elucidate the mechanism underlying  the L-type ferrimagnetism and reproduce the temperature dependence of magnetization using self-consistent mean-field theory.

\begin{figure*}[t]
\centerline{\includegraphics[width=18.cm]{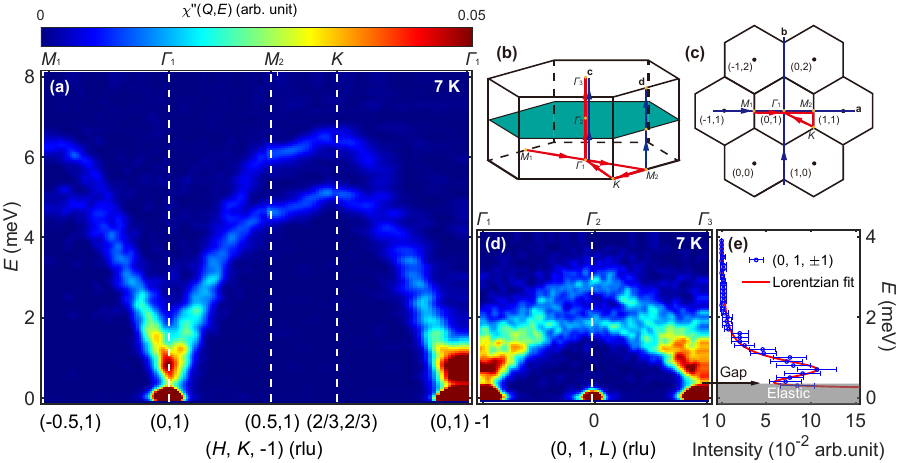}}
\caption{(a) and (d) INS results of the magnetic excitation spectra along high-symmetry paths for the in-plane and out-of-plane directions, respectively. The data were obtained with $E_{\rm i}=12$~meV. In (a), the integration thicknesses of the other orthogonal in-plane direction and the out-of-plane direction are $\pm 0.1$ and $\pm 0.2$~rlu, respectively. In (d), the integration thicknesses of the two in-plane directions are both $\pm 0.1$~rlu. (b) and (c) Schematics of the 3D and 2D reciprocal space of \MMO. Red lines and arrows show the high-symmetry paths of data presented in (a) and (d), while the dark blue lines represent the directions of the dispersions shown in Fig.~\ref{fig3}. (e) Constant-$\bm{Q}$ cuts at $\bm{Q}=(0, 1, \pm 1)$. The data points obtained from $L=\pm1$~rlu are averaged. The arrow indicates the gap position. The solid line is a Lorentzian fit to the data. The shade represents the contamination from the elastic line. (d) and (e) share the same vertical axis.
\label{fig2}} 
\end{figure*}

\begin{table*}[t]
\caption{\label{table1}
The fitted exchange parameters with the unit of meV. Positive/negative $J$ denotes an antiferromagnetic/ferromagnetic exchange interactions. Positive/negative $\Delta$ denotes an easy-axis/easy-plane anisotropy.}
\begin{ruledtabular}
\begin{tabular}{ccccccccc}
			\textrm{$J_{1}$}&\textrm{$J_{2}$}&\textrm{$J_{3}$}/\textrm{$J_{4}$}&\textrm{$J_{1}^{O}$}& \textrm{$J_{2}^{O}$} & \textrm{$J_{1}^{T}$}& \textrm{$J_{2}^{T}$}&\textrm{$\Delta^{O}$}&\textrm{$\Delta^{T}$}\\
			\colrule
			-0.118 & 0.030 & -0.15 & 0.194 & 0.035 & 0.159 & 0.015  & 0.085 & -0.01\\
		\end{tabular}
	\end{ruledtabular}
\end{table*}

\section{Experimental details and sample characterizations}

Single crystals of \MMO were grown using the chemical vapor transport method at temperatures between 945 and 845 $^{o}$C. TeCl$_{4}$ was used as the source of the transport agent~\cite{STROBEL1983329,STROBEL1982242,MCALISTER1983340}. To address the corrosion issue, stoichiometric powders were sealed in double-walled quartz tubes after vacuuming. The resulting as-grown crystals manifested as black hexagonal blocks, with dimensions of approximately 3-4 mm in the $a$-$b$ plane and 1-2 mm along the $c$ axis. {\new The chemical composition of the grown crystals was checked by Energy-dispersive X-ray (EDX) spectroscopy (see details in Appendix A). }

Magnetization measurements were performed using a Physical Property Measurement System (PPMS-9T) from Quantum Design. A magnetic field of 0.1 T was applied both along and perpendicular to the $c$ axis to measure the out-of-plane and in-plane magnetic susceptibility, respectively. For the INS experiments, 180 pieces of single crystals weighed about 3.48~g in total, were coaligned using a back-scattering Laue x-ray diffractometer. The INS experiment was carried out on a direct geometry chopper spectrometer MERLIN at the ISIS Facility in the United Kingdom. A primary energy of $E_{\rm i} = 12.00$ meV and a chopper frequency of 300 Hz were utilized for the measurements, resulting in an energy resolution of 0.34 meV at the elastic line. All measurements were conducted at 7~K. Raw data were processed and analyzed using the Mantid and Horace softwares~\cite{ARNOLD2014156,EWINGS2016132}. The wave vector $\bm Q$ was expressed by $(H,\,K,\,L)$ in reciprocal lattice unit (rlu) of $(a^{*},\,b^{*},\,c^{*})=(4\pi/\sqrt{3}a,\,4\pi/\sqrt{3}b,\,2\pi/c)$ with $a=b=5.80$~{\AA} and $c=10.24$~{\AA}. The measured neutron scattering intensities were corrected by dividing the square of the magnetic form factor and the Bose statistic factor. Theoretical simulations of the excitation spectra were conducted using the SPINW program~\cite{Toth_2015}.

The results of magnetic susceptibility measurements are depicted in Fig.~\ref{fig1}(d). Upon cooling, the out-of-plane magnetic susceptibility exhibits a sudden increase at the transition temperature $T_{\rm C}\sim41.5$~K, indicating the onset of ferrimagnetic order. It reaches a maximum at approximately 35~K and steadily decreases to zero with further cooling. The in-plane magnetic susceptibility follows a similar trend, albeit with much smaller net magnetizations, suggesting the presence of an easy-axis magnetic anisotropy along the $c$ direction. These unusual thermal behaviors of the magnetic susceptibility are reminiscent of those observed in an L-type ferrimagnet~\cite{neel1948proprietes,Long1984,doi:10.1021/ic402616r,PhysRevMaterials.6.094412}, where two sublattices of magnetic ions have moments that are equal in magnitude but opposite in direction. This results in the net magnetic moment being zero at a low temperatures, but unlike in antiferromagnetism, the magnetic moments on the two sublattices are not perfectly balanced across the entire temperature range. To roughly evaluate the magnetic interactions, a Curie-Weiss fit was performed with the magnetic susceptibility data above 150~K, as shown in the inset of Fig.~\ref{fig1}(d). The extracted Curie-Weiss temperatures are $\Theta_{\parallel}=-150.25$ K and $\Theta_{\perp}=-129.18$ K, suggesting antiferromagnetic couplings between the spins in the octahedral and tetrahedral sites~\cite{MCALISTER1983340,PhysRevB.95.045142,10.1063/1.333657}. Furthermore, the effective magnetic moments were derived to be $6.34\ \mu_{\rm B}$ and $6.16\ \mu_{\rm B}$ for $H|| c$ and $H\perp c$, respectively. Considering the Mn$^{2+}$ ion holds an $S=5/2$ high spin state, the resulting Land\'e $g$ factors are $g_{\parallel}=2.14$ and $g_{\perp}=2.08$, respectively, manifesting a weak spin-orbital coupling. For a more comprehensive understanding of the spin interactions, INS measurements were conducted.

\begin{figure}[htb]
\centerline{\includegraphics[width=8.cm]{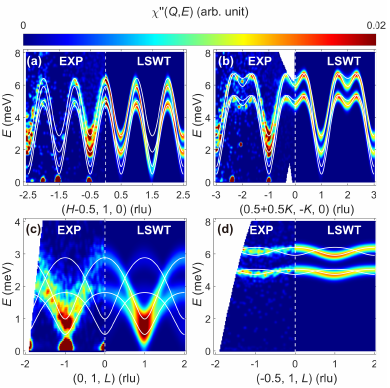}}
\caption{Comparisons between the measured (EXP, left side in each panel) and calculated magnetic excitation spectra using the linear-spin-wave theory (LSWT, right). Directions of the dispersions are depicted in Fig.~\ref{fig2}(b) and (c). The experimental data were collected at 7~K. Solid curves denote the calculated magnon dispersions.
\label{fig3}}
\end{figure}

\section{Results}
Figures~\ref{fig2}(a) and \ref{fig2}(d) show the magnon spectra of \MMO along high-symmetry directions as depicted in Fig.~\ref{fig2}(b) and~\ref{fig2}(c). Overall, two well-defined magnon branches are observed. For the in-plane direction [{Fig.~\ref{fig2}(a)], both magnon bands are found to originate from the zone center $\rm \Gamma$ point and propagate towards the zone boundary. Near the $\rm \Gamma$ point, these two magnon bands nearly overlap, making it challenge to distinguish them. As they propagate towards the zone boundary, these two bands gradually separate and become distinct at the Brillouin zone boundary. The maximum energies of these two magnon bands are 6.8 and 5.4~meV, respectively, both occurring at the K point. The observation of a band top at 6.8~meV in our INS measurements aligns with the results from a previous THz study~\cite{PhysRevB.102.144410}. For the out-of-plane direction~[{Fig.~\ref{fig2}(d)], the magnon excitations exhibit an $L=2$ period, consistent with the two-layer structure of the primitive cell. Additionally, the overlapped magnon bands also separate with maximum energies of 3 and 1.9~meV at the second Brillouin zone center with $L=0$~rlu. To inspect the excitation gap, constant-$\bm{Q}$ cuts at $\bm{Q}=(0, 1, \pm1)$ are taken and the results are shown in Fig.~\ref{fig2}(e). The magnon excitations exhibit a peak at 0.8~meV and a dip at 0.48~meV~[Fig.~\ref{fig2}(e)], indicating the presence of an excitation gap with the value of $\sim$0.48~meV, close to that of 0.5~meV reported in Ref.~\cite{PhysRevB.102.144410} by a THz spectroscopy measurement.

Based on symmetry analysis, we employ the following spin Hamiltonian to model our data:
\begin{equation}\label{H}
\begin{aligned}
H&=-\sum_{\langle i,j \rangle} J_{i,j}\bm{S}_{i}\cdot\bm{S'}_{j}\\&-\sum_{\langle i,j \rangle\in \rm O} J_{i,j}^{\rm O}\bm{S}_{i}\cdot\bm{S}_{j}-\sum_{\langle i,j \rangle\in \rm T} J_{i,j}^{\rm T}\bm{S'}_{i}\cdot\bm{S'}_{j} \\&-\sum_{\langle i \rangle \in \rm O}\Delta_{i}^{\rm O}(\bm{S}^{z}_{i})^{2}-\sum_{\langle i \rangle \in \rm T}\Delta_{i}^{\rm T}(\bm{S'}^{z}_{i})^{2}.
\end{aligned}
\end{equation}
Note that the DM interaction with components lying in the $a$-$b$ plane is allowed by the structural symmetry of Mn$_2$Mo$_3$O$_8$, as shown in Fig.~\ref{fig1}(b). However, only the DM interaction with component parallel to the magnetic moment will contribute to the spin wave dynamics~\cite{PhysRevLett.115.147201}. Since this condition is not met in Mn$_2$Mo$_3$O$_8$, the DM interaction is not included in our parameterization. In Hamiltonian~(\ref{H}), $\bm{S}$ and $\bm{S'}$ are the spin operators of the Mn$^{2+}$ ions with a magnitude of 5/2 on octahedral and tetrahedral sites, respectively. $J$ denotes the exchange interaction between an octahedral spin and a tetrahedral spin, while $J^{\rm O}$ and $J^{\rm T}$ denote the exchange interactions between two octahedral and tetrahedral spins, respectively, as depicted in Fig.~\ref{fig1}(b) and \ref{fig1}(c). $\Delta^{\rm O}$ and $\Delta^{\rm T}$ represent the single-ion anisotropies of the octahedral and tetrahedral spin, respectively, which are taken into account due to the presence of a finite spin gap of $\sim$0.48~meV as observed in Fig.~\ref{fig2}(c) and by previous THz measurements~\cite{PhysRevB.102.144410}. We used the SPINW package to model our data with the linear-spin-wave theory (LSWT)~\cite{Toth_2015}. In our parameterization, we find that when $\Delta^{\rm O}+\Delta^{\rm T}\sim0.08$ meV, LSWT can give a spin gap of $\sim$0.48 meV with little distinction in fit quality. To better constrain these parameters, we set $\Delta^{\rm O}$ to be positive and $\Delta^{\rm T}$ to be negative. This choice is based on the findings in Ref.~\cite{PhysRevB.102.144410}, which reveals that the octahedral and tetrahedral spins possess easy-axis and easy-plane anisotropies, respectively. The refined exchange interaction parameters are listed in Table~\ref{table1}. Note that these values are smaller than those reported in Ref.~\cite{PhysRevB.102.144410}, as our study considers a more comprehensive set of exchange pathways, which results in a reduced strength for each individual exchange interaction.

In order to visually compare with experimental results, we extract the value of the dynamical spin correlation function $S^{\alpha \beta}(\bm{Q},\omega)$. $S^{\alpha \beta}(\bm{Q},\omega)$ is related to the scattering cross section as~\cite{squires1996introduction,shirane2002neutron}:
\begin{equation}
	\frac{d^2\sigma}{d\Omega d\omega}\varpropto |\bm{F}(\bm{Q})|^{2}\sum_{\alpha \beta}(\delta_{\alpha \beta}-\frac{Q_{\alpha}Q_{\beta}}{Q^2})S^{\alpha \beta}(\bm{Q},\omega),
\end{equation}
which is directly measured in INS experiments. Furthermore, the extracted $S^{\alpha \beta}(\bm{Q},\omega)$ is convoluted with the instrumental resolution to simulate the magnetic excitation spectra. The calculated dispersions and simulated magnetic excitation spectra are plotted together with the measured magnetic excitation spectra in Fig.~\ref{fig3}. With the exchange parameters listed in Table~\ref{table1}, the calculated spin excitation spectra of \MMO exhibit four magnon branches. These four branches can be divided into two pairs and each pair yields only one observable magnon mode in the excitation spectra. Additionally, these two pairs of magnon branches with higher and lower energies are found to originate from the octahedral and tetrahedral spins, respectively. It can be found that the simulated magnon spectra are in excellent agreement with the experimental results, indicating the accuracy and reliability of our effective spin model in characterizing the spin dynamics of \MMO.

\begin{figure}[htb]
\centerline{\includegraphics[width=7.cm]{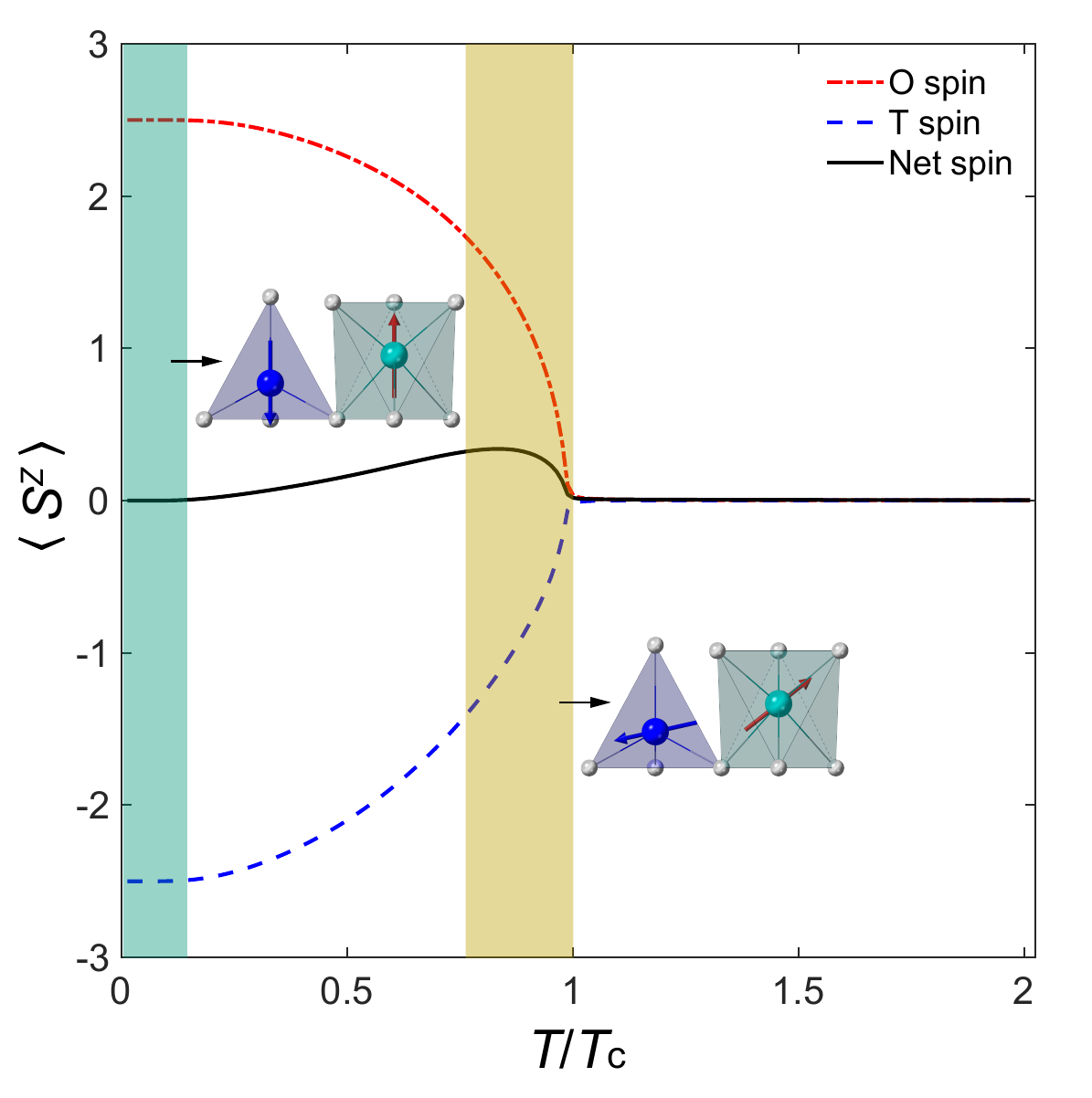}}
\caption{{\new Calculated temperature dependence of the $z$-components of the octahedral, tetrahedral, and net spin moments using self-consistent mean-field theory. The insets are schematic diagrams which show the relative tilting of octahedral and tetrahedral spins from the $c$-axis. Note that the tilting angles just below the transition temperature are arbitrary and only to emphasize that the different tilting angles will result in a net magnetization along the $c$-axis.}}
\label{fig4}
\end{figure}

\section{Discussions and Conclusions}

Apart from different crystal-electric fields of the octahedral and tetrahedral sites ($\Delta^{O} >0> \Delta^{T}$) reported by Ref.~\cite{PhysRevB.102.144410}, another crucial consequence of the bipartite structure revealed by our study is that different exchange paths result in different amplitudes of the spin interactions in the two sublattices ($J_{1}^{\rm O}>J_{1}^{\rm T}>0$, and $J_{2}^{\rm O}>J_{2}^{\rm T}>0$). Notably, the leading exchange interactions in \MMO are the two next-nearest-neighbor interactions $J_{1}^{O}$ and $J_{1}^{T}$, both of which are ferromagnetic. This is different from the other A$_2$Mo$_3$O$_8$ compounds, where the leading exchange term, the nearest-neighbor interaction, is antiferromagnetic~\cite{bao2023direct,PhysRevB.108.024431,gao2023diffusive}. Our derived exchange parameters effectively elucidate the L-type ferrimagnetic magnetization behavior of \MMO. Firstly, it is noteworthy that the half-filled 3$d$ shells of Mn$^{2+}$ ions suggest a quenched orbital moment, resulting in nearly equal magnetic moments at two different sites of the bipartite structure. Considering $J_{1}^{O}$ is the largest interaction and $\Delta^{O}>0$, the phase transition is initiated by the ferromagnetic alignment along $c$ axis of the octahedral spins. Due to the antiferromagnetic interactions ($J_{1}$, $J_{3}$, and $J_{4}$) between octahedral and tetrahedral spins, the alignment of octahedral spins induces an effective molecular field applied to tetrahedral spins, forcing the latter to align ferromagnetically in the opposite direction to the octahedral spins. However, because of $J_{1/2}^{O}>J_{1/2}^{T}$, octahedral spins align faster along the $c$ axis than tetrahedral spins [the inset of Fig.~\ref{fig4}]. Consequently, the spin moments of octahedral and tetrahedral sites do not compensate in the initial stage of the phase transition, resulting in a net magnetization along the $c$ axis. As the Btemperature decreases further, both the spin moments of the octahedral and tetrahedral sites will completely polarize along the $c$ axis, leading to the complete compensation of the net magnetization [the inset of Fig.~\ref{fig4}]. This scenario appears to be justified by a mean-field calculation without detailed exchange parameters~\cite{PhysRevB.95.045142}. To further validate it, we calculated the temperature dependence of octahedral, tetrahedral, and net spin moments by self-consistent mean-field theory using our effective spin model with the exchange parameters listed in Table~\ref{table1} obtained from fitting the INS spectra~\cite{RAVOT1995461,nolting2009quantum} (see details in Appendix B). The results successfully reproduce the trend of the observed L-type ferrimagnetic susceptibility, as depicted in Fig.~\ref{fig4}. The calculated results are scaled by a global factor to directly compare with the measured magnetic susceptibility, which exhibits an excellent match for the ordered state as shown in Fig.~\ref{fig1}(c). The slight mismatch of the background in the paramagnetic phase may be due to the presence of fluctuating moments in the real samples, which is not considered in the calculations. Clearly, the L-type ferrimagnetism of \MMO is a direct consequence of its bipartite structure. 

In summary, our INS study on single crystals of the polar ferrimagnet \MMO has revealed two distinct magnon dispersions, corresponding to the octahedral and tetrahedral spins in \MMO, respectively. The magnetic excitation spectra are effectively described by a spin Hamiltonian encompassing Heisenberg exchange interactions and single-ion anisotropy terms. Our derived effective spin Hamiltonian highlights the distinctive dynamic properties exhibited by tetrahedral and octahedral spins. Furthermore, employing this effective spin model, we successfully replicate the unconventional temperature dependence of the L-type ferrimagnetic susceptibility through self-consistent mean-field theory. Our study not only enhances the comprehension of the spin dynamics inherent in L-type ferrimagnets but also underscores the significance of the bipartite structure in determining the excitation properties of the polar magnets $\rm{A_{2}Mo_{3}O_{8}}$.

\begin{acknowledgments}
We thank Li-Wei He for stimulating discussions. The work was supported by National Key Projects for Research and Development of China with Grant Nos.~2021YFA1400400 and 2024YFA1409200, National Natural Science Foundation of China with Grant Nos.~12225407, 12434005,12074174 and 12404173, Natural Science Foundation of Jiangsu province with Grant Nos.~BK20241250 and BK20241251, Postdoctoral Fellowship Program of CPSF under Grant No.~BX20240161, China Postdoctoral Science Foundation with Grant No.~2024M751367, Jiangsu Funding Program for Excellent Postdoctoral Talent No.~2024ZB021, and Natural Science Foundation of the Higher Education Institutions of Jiangsu Province with Grant No.23KJB140012. We acknowledge the neutron beam time from ISIS with Proposal Nos.~2220304~(DOI: 10.5286/ISIS.E.RB2220304).
\end{acknowledgments}
	
\appendix
{\new 
\section{Composition analyses}
Measurements using EDX spectroscopy were performed to analyze the chemical composition of the grown single crystals. The results are presented in Fig.~\ref{fig5} and Table.~\ref{table2}. The EDX analysis confirms that the main elemental composition of our samples aligns well with the nominal stoichiometry, with minor contributions from unavoidable elements such as carbon and nitrogen. Excluding these minor elements, the atomic percentages obtained from EDX measurements are Mn: 12.5$\%$, Mo: 22.5$\%$, and O: 65$\%$, which are in good agreement with the nominal values of Mn: 15$\%$, Mo: 23$\%$, and O: 61$\%$.

\begin{figure}[htb]
	\centerline{\includegraphics[width=8.cm]{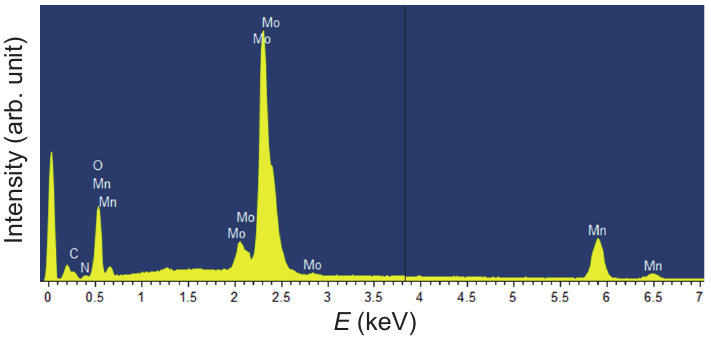}}
	\caption{{\new EDX spectra. Elements are marked on top of the corresponding peaks.}}
	
	\label{fig5}
\end{figure}

\begin{table}[b]
	\caption{\label{table2}
		{\new Element contents obtained from the EDX spectra}}
	\begin{ruledtabular}
		\begin{tabular}{ccc}
			Element & Mass percent & Atomic percent \\
			Mo & 51.5(3) & 18.1(1)\\
			O & 24.6(3) & 52.0(3)\\
			Mn & 16.3(3) & 10.0(4)\\
			C & 4.1(5) & 11.4(2)\\
			N & 3.5(3) & 8.5(1)\\
			  &   &  \\
			Total & 100.00 & 100.00 
		\end{tabular}
	\end{ruledtabular}
\end{table}
}

\section{Calculation of the temperature dependence of the spin moments through self-consistent mean-field theory} 
To reproduce the temperature dependence of the magnetic susceptibility, we begin by the mean-field approximation as follows:
\begin{equation}
	\bm{S}_i\cdot\bm{S}_j \longrightarrow S^z_i\langle S^z_j\rangle + \langle S^z_i\rangle S^z_j-\langle S^z_i\rangle\langle S^z_j\rangle
\end{equation}
where $\langle S^z \rangle$ presents the quantum thermodynamic average of the z-component spin operator $S^z$. The terms with $\langle...\rangle\langle...\rangle$ are neglected, as they contribute insignificantly to the results. A homogeneous mean field is considered, where translational symmetry is present:
\begin{equation}
	\langle S_i^z \rangle=\langle S^z \rangle \forall i
\end{equation}
The effective spin Hamiltonian is then divided into two parts:
\begin{equation}
	\begin{gathered}
		H=H_O + H_T\\
		H_O=-N[3(J_1+J_2)+(J_3+J_4)]\langle S_t^z \rangle S_o^z \\- 2N[3(J_1^O+J_2^O)+D^O]\langle S_o^z \rangle S_o^z-kBS_o^z\\
		H_T=-N[3(J_1+J_2)+(J_3+J_4)]\langle S_o^z \rangle S_t^z \\- 2N[3(J_1^T+J_2^T)+D^T]\langle S_t^z \rangle S_t^z-kBS_t^z
	\end{gathered}
\end{equation}
where the term with the factor $-kB$ is the Zeeman term, and $S_o^z$/$S_t^z$, $H_O$/$H_T$ represent the z-component spin and effective Hamiltonian of octahedral/tetrahedral sublattice, respectively. Let $\phi^{o/t}_{n}$ denote the eigenstates of $H_O$/$H_T$ with $n\in\{-5/2,-3/2,-1/2,1/2,3/2,5/2\}$. The eigenvalues $E^{o/t}_{n}$ of $H_O$/$H_T$ can be solved out using
\begin{equation}
	S_{o/t}^z|\phi^{o/t}_{n}\rangle=n|\phi^{o/t}_{n}\rangle
\end{equation}
The statistical average of $\langle S_{o/t}^z \rangle$ is given by
\begin{equation}
	\langle S_{o/t}^z \rangle=\frac{\sum\limits_{n}\langle \phi^{o/t}_{n}|S_{o/t}^z|\phi^{o/t}_{n}\rangle \mathrm{exp}(-E^{o/t}_{n}/k_{B}T)}{\sum\limits_{n}\mathrm{exp}(-E^{o/t}_{n}/k_{B}T)}\\
\end{equation}
By following above procedure, two self-consistent functions are obtained. The temperature dependence of $\langle S^{z}_{o}\rangle$ and $\langle S^{z}_{t}\rangle$ are acquired by solving these two self-consistent functions through an iterative algorithm. The net spin moments are calculated as
\begin{equation}
	\langle S_n^z \rangle =\langle S_o^z \rangle + \langle S_t^z \rangle.
\end{equation}

%

\end{document}